\documentclass{article}

\usepackage{amsmath}
\usepackage{graphicx}

\begin{document}

\begin{center}
{\large \bf A SIMPLE SOLID-ON-SOLID MODEL OF EPITAXIAL FILM GROWTH: SUBMONOLAYER SUBSTRATE COVERAGE}

\bigskip

{\bf K. Malarz}

\bigskip

{\it Department of Theoretical and Computational Physics,
Faculty of Physics and Nuclear Techniques,
University of Mining and Metallurgy (AGH).\\
al. Mickiewicza 30, PL-30059 Krak\'ow, Poland.}\\

\bigskip

{\it E-mail: malarz@agh.edu.pl}

\bigskip

\today
\end{center}

\begin{abstract}
In this work we investigate the influence of substrate temperature on the surface morphology for substrate coverage below one monolayer.
The model of film growth is based on random deposition enriched by limited surface diffusion.
Also anisotropy in the growth is involved.
We found from computer simulations for the simple cubic lattice and solid-on-solid model, that the surface morphology changes with increasing temperature from isotropically distributed isolated small islands, through anisotropic 1-D stripes to larger 2-D anisotropic islands and again randomly distributed single atoms.
The transition is also marked in height-height correlation function dependence on temperature, as directly seen by snapshots from simulations.
The results are in good qualitative agreement with already published results of kinetic Monte Carlo simulations, as well as with some experimental data.
\end{abstract}


\section{Introduction}
Surface growth is a problem which may be found in mathematics, physics,
biology or even social phenomena, if we define the surface as a boundary
between different phases in $d$-dimensional space \cite{herrmann86,meakin93,family90,levi97}.
Theoretical studies of the properties of the surfaces of the grown
crystal/thin films based on Monte Carlo (MC) simulations, initially using
the solid-on-solid (SOS) models, then extended to more realistic kinetic MC
methods, became particularly popular in early 80's (see \cite{levi97} for
review).
These studies resulted in at least a semi-quantitative understanding of
growth modes and step dynamics.
Then rapid progress in observational methods (including diffractional
methods --- RHEED, LEED and direct imaging --- STM, AFM) allowed for better
understanding mechanism and physics of crystal/films growth.
However, MC simulations may be still applied as a complementary method
which is sometimes more efficient while analytical treatment fails and surely
is more cheaper than performing experiments.

In this work with the SOS model \cite{maksymowicz96,malarz99_1,malarz99_2,malarz00_1,malarz00_2}
we would like to check how substrate temperatures influence qualitatively the
surface morphology before the percolation limit, when only a small fraction of
the first monolayer (ML) is deposited.

\section{Model}
The present model \cite{maksymowicz96,malarz99_1,malarz99_2,malarz00_1,malarz00_2} on a square lattice is based on
random deposition (RD) enriched by a relaxation process of incoming at the
fixed rate particles.
After a random choice of the place of the particle's initial contact to the
surface, particles migrate to one of the four nearest neighbor (NN)
sites or stick on the place of deposition.
The migration procedure is repeated $L_{\text{dif}}$ times, thus
$L_{\text{dif}}$  may be considered a range of diffusion \cite{malarz99_1}.
In the local relaxation process, the particle tends to maximize the number
of particle-particle lateral bonds (PPLB) similarly to the Wolf--Villain model
\cite{wolf90}.
The probability of choosing one of the five accessible sites is given by the
Boltzmann factor $\exp(-E/k_BT)$, where the particle's total energy
$E=n_xJ_x+n_yJ_y$ in all five positions depends on the number $n_x$ ($n_y$)
of the virtually created bonds in $x$ ($y$) direction.
$J_x$ and $J_y$ reflect strengths of the PPLB.
$k_B$ is the Boltzmann constant and $T$ denotes the substrate absolute
temperature.
Additionally, the system tendency to increase the number of PPLB is slowed
down by the barrier for diffusion $V$: the probability of movement (choosing NN
sites) is reduced by a factor $\exp(V_x/k_BT)$ or $\exp(V_y/k_BT)$, depending
on the considered direction of diffusion.
The diffusion barrier $V$ must be positive, while the negative $E$ is compatible
with the assumed tendency of the system to maximize the number of PPLB.

The model control parameters were taken after Ref. \cite{mottet98} (where
kinetic MC simulations were performed) to mimic the realistic ratio of
$V_x/V_y$ and $J_x/J_y$.
We assume that activation energy for diffusion and activation energy for
breaking PPLB (in terminology of the cited paper) may correspond to the
model control parameters $V$ and $-E$, respectively.
$x$-direction is parallel to dimer rows, while $y$ shows the direction of
cross-channel diffusion on a bcc(110) plane:
\[
\begin{split}
\text{Ag:~} \qquad V_x/V_y=0.736 \qquad J_x/J_y=9.000 \\
\text{Cu:~} \qquad V_x/V_y=0.793 \qquad J_x/J_y=6.857
\end{split}
\]

\section{Results of simulations}
The simulations were carried out on a $256\times 256$ lattice with periodic
boundary conditions.
Diffusion was limited to $L_{\text{dif}}=10^3$ steps.
We measure the surface anisotropy in terms of the dimensionless parameter
$\varepsilon\equiv \left[ G(1,0)-G(0,1)\right] / G(0,0)$,
where $G(\mathbf{s})\equiv \langle h(\mathbf{r}+\mathbf{s})\cdot
h(\mathbf{r}) \rangle - \langle h(\mathbf{r}) \rangle ^2$
is height-height correlation function \cite{malarz99_2}.
The average $\langle\ldots\rangle$ runs over all substrate sites
$\mathbf{r}$ and the dispersion $G(0,0)=\sigma^2=\langle\left[ h(\mathbf{r})-
\langle h(\mathbf{r})\rangle\right]^2 \rangle$
is defined as a square of the surface width $\sigma$, which usually describes
the surface roughness.

\subsection{Influence of the substrate temperature on the surface 
anisotropy}
\begin{figure}
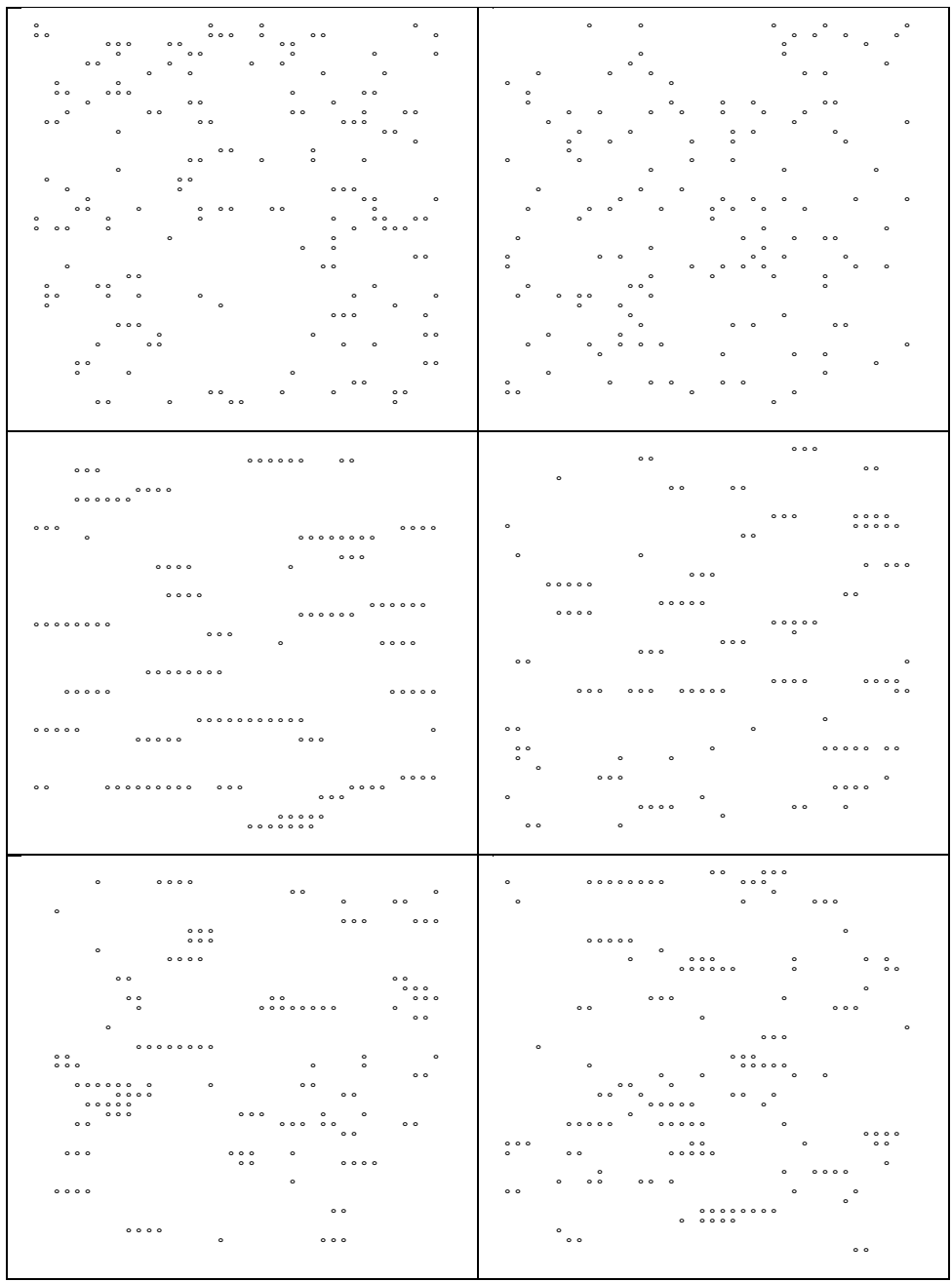

\begin{center}
\begin{tabular}{|c|c|}
\hline
\includegraphics[width=6cm]{malarz01.fig}&
\includegraphics[width=6cm]{malarz02.fig}\\ \hline
\includegraphics[width=6cm]{malarz03.fig}&
\includegraphics[width=6cm]{malarz04.fig}\\ \hline
\includegraphics[width=6cm]{malarz05.fig}&
\includegraphics[width=6cm]{malarz06.fig}\\ \hline
\end{tabular}
\caption{Snapshots from simulations for coverage $\theta=0.1$ ML for a
$40\times 40$ cut out net of the substrate.
Dots correspond to occupied sites.
Substrate temperature grows from top to bottom of figure.
Columns correspond to Ag-like and to Cu-like parameters, respectively.}
\label{fig_snapshots}
\end{center}
\end{figure}

\begin{figure}
\begin{center}
\includegraphics[width=6cm,angle=-90]{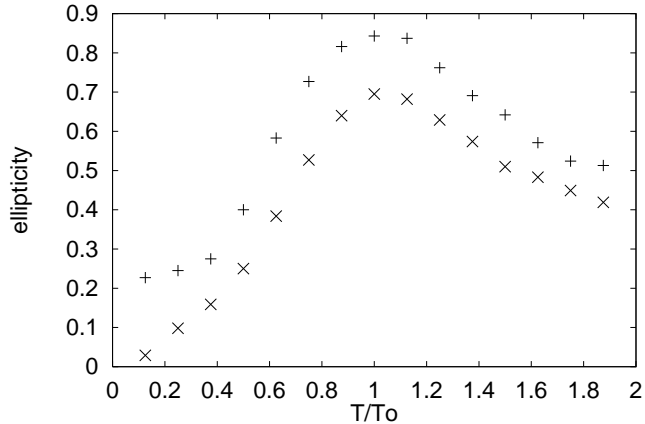}
\caption{Anisotropy parameter $\varepsilon$ for different substrate
temperatures $T$ (normalized to the temperature $T_0$ for which $\varepsilon$
is maximal) for homoepitaxial growth on (110) plane for Cu-like ($\times$)
and Ag-like ($+$) model control parameters.
Net size is $256\times 256$.
The substrate is covered by $\theta=0.1$ ML.}
\label{fig_epsilon}
\end{center}
\end{figure}

Fig. \ref{fig_snapshots} shows changes of surface morphology with increasing
temperature for submonolayer coverage $\theta=0.1$~ML.
For the limiting case $T\to 0$ ($\exp(V/k_BT)\to\infty$), the large diffusion
barriers stop any migration.
Thus the probability to stay in the place of deposition $P_0$ goes to unity,
while the
probability of the migration to the NN sites $P_{\text{NN}}$ goes to zero, 
and a film
surface grown on {\em a very cold} substrate contains randomly, isotropically
distributed single atoms.
The anisotropy parameter $\varepsilon$ tends to zero.
Increase of the temperature changes the ratio between Boltzmann factors
drastically.
For {\em low} temperatures isotropically distributed small aggregates on the
substrate appear (see the upper part of Fig. \ref{fig_snapshots}).
In such a case although particles are allowed to make $L_{\text{dif}}=10^3$
steps, they stick after making only a few of them, and  the surface
morphology still corresponds roughly to that generated with RD rules.
Then in {\em medium} temperatures range --- see the medium part of
Fig. \ref{fig_snapshots} --- migration in $y$-direction is still stopped by
diffusion barriers, while particles are able to diffuse in $x$-direction
and long 1-D stripes on the substrate grow.
The anisotropy parameter $\varepsilon$ becomes maximal.
Finally, in the bottom of Fig. \ref{fig_snapshots} nucleation of larger
islands is observed with tendency of their orientation along $x$-direction
when the temperature is relatively {\em high}.
For infinitely large temperature $T\to\infty$ ($\exp(V/k_BT)\to 1$), the
barrier for diffusion does not influence the particle migration.
The probabilities of choosing any of the accessible sites become equal
($P_0=P_{\text{NN}}=1/5$).
Particles migration is stopped only by limited to $L_{\text{dif}}$ steps
diffusion.
Thus, for {\em very hot} substrates the surface morphology is similar to that
generated with RD and again $\varepsilon=0$.
The changes of the surface morphology may also be observed by the temperature
dependence of the height-height correlation function $G(\mathbf{s})$.
In Fig. \ref{fig_epsilon}, as discussed earlier, the influence of the
substrate temperature $T$ on the surface morphology anisotropy parameter
$\varepsilon$ is presented.

\subsection{Influence of the substrate temperature on the surface roughness}
For submonolayer coverage, the film contains $\theta L^2$ occupied substrate
sites (with $h=1$) and $(1-\theta)L^2$ empty ones ($h=0$).
Neither infinitely large diffusion $L_{\text{dif}}\to\infty$ nor varying
substrate temperature $T$ can change the distribution of film heights and/or
influence the single-site characteristic $\sigma$ --- the surface roughness
(they influence, however, two-sites characteristics such as $\varepsilon$ --- the
surface anisotropy).
Thus roughness $\sigma$ grows like $\sqrt{\theta}$ as it is predicted by
Poisson process for RD case.
For example, for surfaces which exhibit self-affinity and, as a results of
that, their roughness dynamics may be described by Family--Vicsek law
\cite{family85},
before completing first ML --- or more precisely, before substrate coverage
reach the percolation threshold and the notion of film thickness can be
applied --- surface roughness dynamics corresponds to simple RD.
Such a situation was already observed for many SOS models, and explained as
an artifact of an initially flat substrate \cite{malarz99_1,family86,biehl98,dassarma91}.
For instance, in the Family model \cite{family86}, particles after RD perform
a local search for the column with minimal height.
Thus, starting with perfectly flat substrate, and for submonolayer coverage,
all sites are energetically equivalent and diffusion is unimportant.
Note, that for submonolayer MBE also other law to describe dynamic scaling
was proposed \cite{amar94}.

\section{Conclusions}
In summary, we found from computer simulations that with increasing substrate
temperature a transition in surface morphology takes place.
In contrast, varying the substrate temperature does not influence the film roughness
for submonolayer coverage.
The changes in surface morphology are due to changes of the relative ratio
between Boltzmann factors for diffusion barriers and strengths of attraction
(bonds energies) in two directions perpendicular to each other.

Qualitative agreement with kinetic MC results of simulation with fully
reversible aggregation SOS model for Ag(110) and Cu(110) \cite{mottet98,ferrando97,ferrando98} and
experimental observation for Cu submonolayers \cite{roder93,hahn94} was achieved.
Also STM images \cite{mo91} and hot STM movies \cite{pearson96} of
homoepitaxial submonolayer growth on Si(001) substrate shows similar
morphology transition.
For low deposition rates, 1-D islands grow well within a 50~K window around 530~K.
At lower temperatures, mobility is decreased, and atoms deposited on the
surface nucleate rather new islands than attaching to existing islands.
At higher temperatures islands merge and become 2-D but remain still
anisotropic \cite{pearson96}.
At low coverage and at room temperature III group elements (Ga, In) and IV
group elements (Si, Ge, Sn, Pb) all form 1-D chains of ad-dimers \cite{shen97}.
Shen and co-workers --- as it was earlier stated also in the work of Mo et al. 
\cite{mo89} and confirmed by STM studies \cite{pearson96} --- conclude that,
for such a systems, not only the anisotropy in the diffusion but also the
anisotropy in the sticking must be involved in modeling film growth.
In presented here model first assumption is realized by different barriers for
diffusion ($V_x\ne V_y$) while
anisotropy in bonds ($J_x\ne J_y$) mimics anisotropic sticking probability.

\section*{Acknowledgments}
I am grateful to A.Z.~Maksymowicz for scientific guidance and fruitful discussion.
The simulations were carried out in ACC-CYFRONET-AGH.
This work and machine time in ACC-CYFRONET-AGH are financed by Polish
Committee for Scientific Research (KBN) with grants No. 8~T11F~02616 and
No. KBN/\-S2000/\-AGH/\-069/\-1998, respectively.




\begin{thebibliography}{99}
\bibitem{herrmann86}
H.J. Herrmann,
{\em Phys. Rep.} {\bf 136} (1986) 153.
\bibitem{meakin93}
P. Meakin,
{\em Phys. Rep.} {\bf 235} (1993) 189.
\bibitem{family90}
F. Family,
{\em Physica} {\bf A168} (1990) 561.
\bibitem{levi97}
A.C. Levi, M. Kotrla,
{\em J. Phys.: Condens. Matter} {\bf 9} (1997) 299.
\bibitem{maksymowicz96}
A.Z. Maksymowicz, M.S. Magdo\'n, J.S.S. Whiting,
{\em Comp. Phys. Comm.} {\bf 97} (1996) 101.
\bibitem{malarz99_1}
K. Malarz, A.Z. Maksymowicz,
{\em Int. J. Mod. Phys.} {\bf C10} (1999) 645.
\bibitem{malarz99_2}
K. Malarz, A.Z. Maksymowicz,
{\em Int. J. Mod. Phys.} {\bf C10} (1999) 659.
\bibitem{malarz00_1}
K. Malarz, A.Z. Maksymowicz,
{\em Thin Solid Films} {\bf 367} (2000) 28.
\bibitem{malarz00_2}
K. Malarz,
{\em Electron Technology} {\bf 33} (2000) 319.
\bibitem{wolf90}
D.E. Wolf, J. Villain,
{\em Europhys. Lett.} {\bf 13} (1990) 389.
\bibitem{mottet98}
C. Mottet, R. Ferrando, F. Hontinfinde, A.C. Levi,
{\em Surf. Sci.} {\bf 417} (1998) 220.
\bibitem{ferrando97}
R. Ferrando, F. Hontinfinde, A.C. Levi,
{\em Phys. Rev.} {\bf B56} (1997) 4406.
\bibitem{ferrando98}
R. Ferrando, F. Hontinfinde, A.C. Levi,
{\em Surf. Sci.} {\bf 402-404} (1998) 286.
\bibitem{family85}
F. Family, T. Vicsek,
{\em J. Phys.} {\bf A18} (1985) L75.
\bibitem{family86}
F. Family,
{\em J. Phys.} {\bf A19} (1986) L441.
\bibitem{biehl98}
M. Biehl, W. Kinzel, S.Schinzer,
{\em Europhys. Lett.} {\bf 41} (1998) 443.
\bibitem{dassarma91}
S.~Das Sarma, P. Tamborenea,
{\em Phys. Rev. Lett.} {\bf 66} (1991) 325.
\bibitem{amar94}
J.G. Amar, F. Family, P.-M. Lam,
{\em Phys. Rev.} {\bf B50} (1994) 8781.
\bibitem{roder93}
H. R\"oder, E. Hahn, H. Brune, J.P. Bucher, K. Kern,
{\em Nature} {\bf 366} (1993) 141.
\bibitem{hahn94}
E. Hahn, E. Kampshoff, A. Fricke, J.P. Bucher, K. Kern,
{\em Surf. Sci.} {\bf 319} (1994) 277.
\bibitem{mo91}
Y.-W. Mo, J. Kleiner, M.B. Webb, M.G. Lagally,
{\em Phys. Rev. Lett.} {\bf 66} (1991) 1998.
\bibitem{pearson96}
C. Pearson, M. Krueger, E. Ganz,
{\em Phys. Rev. Lett.} {\bf 76} (1996) 2306.
\bibitem{shen97}
T.-C. Shen, C. Wang, J.R. Tucker,
{\em Phys. Rev. Lett.} {\bf 78} (1997) 1271.
\bibitem{mo89}
Y.-W. Mo, B.S. Swartzentruber, R. Kariotis, M.B. Webb, M.G. Lagally,
{\em Phys. Rev. Lett.} {\bf 63} (1989) 2393.
\end{thebibliography}
\end{document}